\documentclass[%
 reprint,
nofootinbib,
 amsmath,amssymb,
 aps,
 pra,
floatfix,
]{revtex4-1}
\usepackage{comment}
\usepackage{graphicx}
\usepackage{dcolumn}
\usepackage{bm}
\usepackage[table]{xcolor}
\usepackage{booktabs}
\usepackage[toc,page]{appendix}
\usepackage{lipsum}
\usepackage[symbol]{footmisc}

\newcommand{\bra}[1]{\langle#1|}
\newcommand{\ket}[1]{|#1\rangle}
\newcommand{\braket}[2] {\langle#1|#2\rangle}
\newcommand{\Schro}{Schr\"{o}dinger}

\begin{document}

\preprint{APS/123-QED}

\title{A Comparison of Numerical Approaches to the Solution of the Time-Dependent {\Schro} Equation in One Dimension}

\author{H. Gharibnejad}
\author{B. I. Schneider}%
\affiliation{%
National Institute of Standards and Technology, Gaithersburg, Maryland 20899
}%

\author{M. Leadingham}
\affiliation{
 West Virginia
Wesleyan College, Buckhannon, West Virginia 26201
}%

\author{H. J. Schmale}
\affiliation{
 Millersville University, Millersville, Pennsylvania 17551
}%

\date{\today}

\begin{abstract}
We present a simple, one-dimensional model of an atom exposed to a time-dependent intense, short-pulse EM field with the objective of teaching undergraduates how to apply various numerical methods to study the behavior of this system as it evolves in time using several time propagation schemes.

In this model, the exact coulomb potential is replaced by a soft-core interaction to avoid the singularity at the origin. While the model has some drawbacks, it has been shown to be a reasonable representation of what occurs in the fully three-dimensional hydrogen atom. 

The model can be used as a tool to train undergraduate physics majors in the art of computation and software development.



\end{abstract}

\pacs{Valid PACS appear here}
\maketitle


\section{\label{sec:level1}Introduction}

Simulation and data analytics have become increasingly important to all areas of physics as problems become more complex and less amenable to analytic solutions. Advanced undergraduates in physics would greatly benefit from solving a realistic physical problem and one that exposes them to some of the computational techniques that physicists use to solve such problems from scratch. This helps the students
to simultaneously learn numerical analysis and hone their programming skills.

While high level applications, such as Matlab and Mathematica, provide students with useful computational toolkits, it is often the case that students use these as ``black boxes", having little understanding of what is really going on ``under the hood". Furthermore, students learning experience could be greatly enhanced if they study and implement numerical methods in the context of a physical simulation.


This study is the result of collaboration of H.G. and B.I.S with two  undergraduates M.L and H.J.S., who were participating in NIST's summer internship program, SURF.    
 Neither students had any experience with finite difference methods or the solution of differential equations. Both received training to understand the quantum mechanics underlying the study and to guide them in programming the model. The senior investigators helped the students in implementing the codes in Fortran.
 The students were quick in picking up the skills required to produce code which was both bug-free and efficient. 

The goal of this study is to examine a one-dimensional model of the hydrogen atom exposed to a strong, time dependent electromagnetic (EM) field. In this model, the exact coulomb potential is replaced by a soft-core interaction to avoid the singularity at the core. This model has been the subject of numerous papers in the literature~\cite{loudon_one-dimensional_1959,haines_one-dimensional_1969,j._javanainen_numerical_1988,loudon_one-dimensional_2016} and has been shown to be a reasonable representation of what occurs in the fully three-dimensional hydrogen atom.  While similar time independent models have been the subject of many pedagogical studies \cite{merrill_computer_1971,greenhow_continuum_1992,chau_when_1995}, atoms exposed to a time-dependent, intense, short-pulse EM field are of significant current interest \cite{drescher_time-resolved_2002,uiberacker_attosecond_2007,eckle_attosecond_2008} and the performance of various time-propagation methods are a subject of renewed attention in the theoretical AMO community. 


Several of the time-propagation schemes we discuss in this paper have been compared in a classic paper by Leforestier et. al. \cite{leforestier_comparison_1991}, using a spectral method. Here we present a detailed  
comparison of a number of time-propagation methods in the context of a problem employing a finite-difference method. The goal is to have a more accessible and easily programmable set of instructions for students. For completeness, the senior authors used a few higher order finite difference schemes to demonstrate how these can dramatically increase accuracy and performance.  

The paper contains nine sections and two appendices.  The appendices were added to give the reader some foundational information on the time-dependent {\Schro} equation (TDSE) and the short-time approximation to evolving the {\Schro} equation in time.

The paper is organized as follows:
Section~\ref{sec:TDSE} discusses the fundamental assumptions for the TDSE. This section is complemented with the lengthier  derivations of Appendices \ref{appsec:TDSE} and \ref{appsec:SolutionsTDSE}.

Section~\ref{sec:gencomments} contains a brief introduction to the classical electromagnetic field (EM) and gauge invariance. 

In section~\ref{sec:genhamiltonian} we discuss the general properties of a quantum mechanical Hamiltonian that involves the interaction of an electromagnetic field (EM) with an atom, emphasizing the question of the choice of gauge. 

In section~\ref{sec:1dmodel}, we  explain the 1-D model of hydrogen atom that we used as a test case for propagating the time-dependent TDSE.
Here, we also detail the length and velocity gauge forms of the model, discuss its discretization scheme, and mention the computational parameters used throughout the paper. 

In section~\ref{sec:compmethods}, we examine the computational methods used for propagating the solutions to the TDSE. 

In section~\ref{sec:data}, the methods are compared to one another in terms of their relative performance. The comparisons and results are shown for  bound-state excitations as well as above threshold ionization (ATI).
The results indicate that while many numerically simple (low order in time)  propagation schemes work, they often require quite small time steps for high accuracy. Comparing them against more accurate methods, requiring more work per time step but allowing larger time steps, is illuminating. 
In cases where comparisons with published results could be made, the agreement was quite good and gave the students confidence in what they were doing.


Finally, we present our conclusions in section~\ref{sec:conclusions}.

\section{\label{sec:TDSE}Time-Dependent Schr{\"o}dinger Equation (TDSE)}
Atomic units ( $e=\hbar=m=1$ ) are used throughout the article.
\footnote[1]{Atomic units are units most appropriate at the scale of an electron, for more details see \cite{foot_summary_2005}}

The time-dependent {\Schro} equation in quantum mechanics is represented by the differential equation (see Appendix \ref{appsec:TDSE})

\begin{equation}
i  \frac{\partial \psi}{\partial t} = \mathbf{H} \psi\,,
\end{equation}
where $\mathbf{H}$ is the Hamiltonian of the system and $\psi$ is the wavefunction. 
\par

As shown in Appendix \ref{sec:solutions}, the solutions to the TDSE for a general time dependent Hamiltonian is quite complex. The usual procedure is to:
\begin{enumerate}
\item Make a short-time approximation to the  time ordered exponential by ignoring the time-ordering for very short times.
\item Introduce some discretization in space via a finite-difference or basis set expansion.
\item Develop a time evolution scheme for what is now a large matrix exponential.
\end{enumerate}
Converging the solution to sufficient accuracy is achieved by increasing the number of spatial basis functions - or grid points - and decreasing the size of the time-step so that the propagated wavefunction is essentially unchanged. Short-time propagation requires the action of the exponentiated matrix on a known vector.
Here we investigate a number of methods that perform this propagation with the goal of quantifying their accuracy and computational efficiency.
 
\section{\label{sec:gencomments}Gauge Considerations}
We begin discussion of gauges by writing Maxwell's equations (in Gaussian units)
\begin{subequations}
\label{eq:Maxwell_EB}
\begin{align}
&\boldsymbol{B} = \nabla \times \boldsymbol{A}\,, \\
\label{eq:Maxwell_E}
&\boldsymbol{E}=-\nabla \phi -\frac{1}{c}\frac{\partial\boldsymbol{A}}{\partial t}\,, \\
\label{eq:Mawell2}
&\nabla^2 \phi +\frac{1}{c}\frac{\partial}{\partial t} \Big( \space \boldsymbol{\nabla} \cdot {\boldsymbol A} \Big ) = -4\pi \rho \,, \\
& \Big( \nabla^2 - \frac{1}{c^2} \frac{\partial^2}{\partial t^2}  \Big) \space {\boldsymbol A}  - \nabla \Big( \space \nabla \cdot {\boldsymbol A} +\frac{1}{c}
\frac{\partial}{\partial t} \phi \Big) = - \ \frac{4\pi}{c}{\boldsymbol J}\,.
\end{align}
\end{subequations}
Here $\boldsymbol{E}$ and $\boldsymbol{B}$ are the EM fields, $\boldsymbol A$ and $\phi$ are the vector and scalar potentials, $\rho$ is the charge density and ${\boldsymbol J}$ is the current density.  

The 
observable EM fields, $\boldsymbol{E}$ and $\boldsymbol{B}$, are said to be gauge invariant under the transformations,
\begin{subequations}
\label{eq:gauge_tran}
\begin{align}
\label{eq:gauge_trans_A}
&\boldsymbol{A} \rightarrow \boldsymbol{A}^\prime = \boldsymbol{A} +\boldsymbol{\nabla} \chi\,, \\
\label{eq:gauge_trans_phi}
&\phi \rightarrow \phi^\prime = \phi-\frac{1}{c}\frac{\partial \chi}{\partial t}\,,
\end{align}
\end{subequations}
where $\chi$ is any twice-differentiable function.  
Gauge invariance can be quite useful in simplifying the mathematical equations.
Here we consider two types of gauge representations for a given electric field of the form
\begin{equation}\label{eq:Electricfield}
\boldsymbol{E} = E_0 \hspace{.05cm} \mathrm{sin}(kz-\omega t) \boldsymbol{\hat x}\,.
\end{equation}
One may choose the scalar and vector potentials to be,
\begin{subequations}
\label{eq:vgpotentials}
\begin{align}
 &\phi =0\,,\\
 \label{eq:vgpotentials2}
 &\boldsymbol{A} = E_0 \hspace{.05cm} \frac{\mathrm{cos}(kz-\omega t)}{k}\boldsymbol{\hat x}\,, \\
 &\boldsymbol{\nabla} \cdot \boldsymbol{A} = 0.
\end{align}
\end{subequations}
Here the electric field is propagating in the z direction and is polarized along the x axis. We used the relation $c/\omega=1/k$ and Eq.(\ref{eq:Mawell2}) to derive Eq.(\ref{eq:vgpotentials2}).
These equations are consistent with~\eqref{eq:Maxwell_EB} and~\eqref{eq:gauge_tran} and represent a particular choice of gauge, called the Coulomb-velocity gauge.

We will now make a gauge transformation by choosing the function
\begin{equation}
\label{eq:dipole_trans}
 \chi = - x \hspace{.05cm}\frac{\mathrm{E_0 \hspace{.05cm} cos}(\omega t)}{k}\,,
\end{equation}
which, in the limit that $kz<<1$, yields 
\begin{subequations}
\label{eq:lgpotentials} 
\begin{align}
 &\phi = - x \mathrm{E_0 \hspace{.05cm} sin}(\omega t) \approx -\boldsymbol{x\cdot E}\,,  \\
&\boldsymbol{A} = E_0 \hspace{.05cm} \frac{\mathrm{cos}(kz-\omega t) - \mathrm{cos}(\omega t)}{k} \boldsymbol{\hat x} \approx 0\,.
\end{align}
\end{subequations}
This second choice is known as the Coulomb-length gauge. In what follows we often refer to these two gauge representations as the velocity and length gauge. We restrict our treatment to these two gauges even though in the dipole approximation regime an \textit{acceleration} gauge is also often used \cite{bandrauk_atoms_2013}.

The argument that $kz<<1$ is justified by noting that in the dipole approximation the size of the atomic system is much smaller than the wavelength of the radiation, so one may replace \(e^{i k \hat{n}\cdot x-i\omega t}\) by \(e^{-i\omega t}\).   
(see, for example, Peskin and Moiseyev\cite{peskin1993}.)

\section{\label{sec:genhamiltonian}General Hamiltonian}

The Hamiltonian for a hydrogen-like atom in an EM field may be written as ~\cite{cohen-tannoudji_quantum_2005}
\begin{subequations}
\label{eq:GeneralHinField}
\begin{align}
\label{eq:GF1}
\mathbf{H}=& \frac{1}{2} \big[ p-\frac{1}{c}\boldsymbol{A}\big]^2 +  V +  \phi\,, \\
\label{eq:GF2}
\mathbf{H} =& \frac{1}{2} p^2 -\frac{1}{2c} [ \boldsymbol{A \cdot p  + p \cdot A} ]  \nonumber \\
  &             +   \frac{A^2}{c^2}  + V +\phi\,,
\end{align}
\end{subequations}
where $V$ is the Coulomb interaction of the proton and the electron, depending on space, and $\phi$ is any scalar field associated with the external EM field, depending on time and space. 
The earliest incarnation of a Hamiltonian of the form in~\eqref{eq:GeneralHinField} appears to go back to Karl Schwarzschild~\cite{schwarzschild1903} in 1903.  He did not consider the issue of EM gauges, let alone quantum mechanics, but here it is worthwhile to ask the questions about the EM gauge being used in writing Eqs.~\eqref{eq:GeneralHinField} and how it might be changed via a gauge transformation.  

For low intensity light( $ \le 10^{15}$ W/cm$^2$ ), $A^2$ in Eq.\eqref{eq:GF2} may be neglected. Care is required to ensure this approximation is valid when dealing with very high intensity laser fields. 

Using the chain-rule one can see 
\begin{equation}
\boldsymbol{p} \cdot \boldsymbol{A}\psi  = (\boldsymbol{p} \cdot \boldsymbol{A})\psi + \boldsymbol{A}\cdot \boldsymbol{p}\psi.
\end{equation}
If we take advantage of the Coulomb gauge condition, $\boldsymbol{\nabla} \cdot \boldsymbol{A} = 0$ as well as $\boldsymbol{p}=-i\boldsymbol{\nabla}$, we can simplify the Hamiltonian of~Eq.(\ref{eq:GF2}) to 
\begin{equation}
\label{eq:SimplifiedHinField}
\mathbf{H} = \frac{1}{2} p^2 -\frac{1}{c}  \boldsymbol{A \cdot p }+ V +\phi\,. 
\end{equation} 
Therefore, Eqs.~\eqref{eq:vgpotentials} and~\eqref{eq:lgpotentials} lead to two different but mathematically equivalent forms of the Hamiltonian; one containing the vector potential and momentum and the other containing the electric field and coordinate . The question of whether they are equivalent numerically is a different matter.  For example, the length form gets quite large in the limit of $x$ going to infinity and could possibly lead to unreliable numerical results.

It has also been found that a gauge which emphasizes parts of an approximate wavefunction may give poor results when the gauge used depends more heavily on the wavefunction in that region of space \cite{bandrauk_atoms_2013, dorr_atomic_1990,cormier_optimal_1996}.  Also, some gauges tend to be more accurate at certain wavelengths or intensities and going beyond the dipole approximation.\cite{selsto_alternative_2007}

For the numerical studies in this paper we have explicitly verified that our results are independent of the two gauges employed in the studies. This gives us confidence in our conclusions and justifies that we have not explicitly examined the acceleration gauge.  We have no reason to believe the results obtained in the acceleration gauge or for that matter any gauge would produce something different from what is in the paper.  In addition, the added numerical complexity does not justify the additional work.

In the next section we look in more detail at the Hamiltonian of the 1-D system we are going to be discussing throughout this article in the length and velocity.

\section{\label{sec:1dmodel}The One dimensional Hydrogen Atom}
\subsection{\label{sec:LengthGauge}Outline in Length Gauge}

The model we use throughout this paper is of a particle in a 1-D attractive potential and under the influence of an external potential  that changes over time. In what follows we always assume that the system is prepared in its ground state. 

The Hamiltonian for the one-dimensional model of the hydrogen atom in an external EM field is given in the length form as, 
\begin{equation} \label{eq:FullHamiltonian}
\boldsymbol{H}(t)=\boldsymbol{H}_0-\boldsymbol{x}\cdot \boldsymbol{E}(t)\sin(\omega t )\,,
\end{equation}
with the time-independent part
\begin{subequations}
\begin{align}\label{eq:H0}
\boldsymbol{H}_0 = -\frac{1}{2} \frac{d^2}{dx^2}+ \boldsymbol{V}(x)\,, \\
\boldsymbol{V}(x)=-\frac{1}{\sqrt{1+x^2}}\,. \label{eq:v(x)}
\end{align}
\end{subequations}
The soft-core potential in Eq.~\eqref{eq:v(x)} is identical to  that used in Refs.  \cite{j._javanainen_numerical_1988,eberly_high-order_1989}.  The model  has both virtues and faults.  On the positive side,it preserves the qualitative aspects of the Rydberg and continuous spectrum of the true hydrogen atom.  This is important in seeing how discrete and above threshold ionization peaks behave. However, its failing is that it does not contain the angular momentum coupling inherent in the 3-D hydrogen atom.

The 1-D model has been widely used in applications ranging from high magnetic field interaction with hydrogen to semiconductor quantum wires, carbon nanotubes, and polymers~\cite{loudon_one-dimensional_2016}.  
In retrospect, the s-wave physical H-atom Hamiltonian, which is defined on the half interval, could also have been chosen, but since we wanted to make contact with the work in~\cite{j._javanainen_numerical_1988}, we did not employ it.

In what follows we consider laser field interaction terms with both a smooth pulse envelope,
\begin{equation}\label{eq:E(t)_smooth}
 E(t)=\begin{cases}
    E_0\sin^2\left[\cfrac{\pi t}{T}\right]\,,& \quad 0\leq t \leq T\\
                                                                  \\
    0              & \quad\text{otherwise}\, ,
\end{cases} 
\end{equation}
as well as a square pulse,
\begin{equation}\label{eq:E(t)_square}
 E(t)=\begin{cases}
    E_0,& \quad 0\leq t \leq T\\
                                                                  \\
    0              & \quad\text{otherwise}\, ,
\end{cases} 
\end{equation}
over a time interval $T$. There is a detailed discussion about the use of each of these laser pulse types in Ref. \cite{j._javanainen_numerical_1988}.

In order to study the convergence of the excitation of low-lying bound states with respect to the size of the time-step for various methods, we employed a smooth pulse.  For the above threshold ionization (ATI) continuous spectrum, we used a square pulse in order to compare with Ref.~\cite{j._javanainen_numerical_1988}. We emphasize that our findings concerning the relative timings of the methods and their convergence with step size are independent of the type of pulse employed.  This will be discussed in more detail in section \ref{sec:data}.

 \subsection{\label{sec:velocityGauge}The Velocity Gauge}
Javanainen et al.~\cite{j._javanainen_numerical_1988} suggest that the dipole length form of the time-dependent interaction could cause problems at large distances.  In reality, the interaction is turned off at $T$, so there is no formal problem.  However, if the computational region is large, as it may need to be in order to study ionization, there could be numerical issues.  

To ensure the accuracy of the numerical results, we have performed most of the calculations in both length and velocity gauge.  We have found no differences in the converged results.  In the the length gauge, the interaction with the external EM field is local, producing a Hamiltonian which is a real operator and that makes the calculation more straightforward.  In the velocity gauge, the Hamiltonian is Hermitian and the interaction is non-local. In more realistic problems, the choice of gauge can make significant differences depending on the wavelength of the radiation.  This is a consequence of the presence of high angular momenta in the coupling of the radiation field to the electrons.  A detailed discussion of this remains outside the scope of the present study, but has been commented on in the literature (see for example, Refs. \cite{dorr_atomic_1990,cormier_optimal_1996,bandrauk_atoms_2013} and references therein).

In the velocity gauge,
for the square pulse ($\phi=0$ ),
\begin{equation}\label{eq:vectorandEfieldsrelation}
A(t)=-c \int_0^t E(t')dt'=-2c E_0\cfrac{\sin^2(\omega t/2)}{\omega}\,,
\end{equation}
and for the smooth pulse, 
\begin{equation}\label{eq:vectorandEfieldsrelationSmooth}
A(t)= c E_0 \left( \frac{\sin^2(a^+ t/2 )}{2a^+}- \frac{\sin^2(a^- t/2 )}{2a^-}- \frac{\sin^2(\omega t/2 )}{\omega}\right)\,,
\end{equation}
where 
\begin{align}
&a^+ = 2 \pi/T + \omega\,, \\
&a^- = 2 \pi/T - \omega\,.
\end{align}

To go from $\psi$ in the length gauge to $\psi'$ in the velocity gauge, we use the transformation,
\begin{equation}\label{eq:relationpsipsiprime}
\psi(x,t)=e^{-ix A(t)}\psi'(x,t).
\end{equation}

\subsection{\label{subsec:discretizinggrid}Discretization}
To solve Eq.\eqref{eq:FullHamiltonian}, we divide space into a set of $(2N+1)$ equidistant points ($x_n = n\delta x$) centered at the origin and apply the lowest order central difference formula to discretize the first and second derivatives of the Hamiltonian.  In the length gauge this yields,
\begin{align}\label{eq:tridiaglength}
(H \psi)_n = -\frac{1}{2(\delta x)^2}(\psi_{n+1}-2\psi_n+\psi_{n-1})&+V(x_n)\psi_n \nonumber \\ 
-x_n E(t) \sin(\omega t + \phi)\psi_n\,,
\end{align}
and in the velocity gauge,
\begin{align}\label{eq:tridiagvelocity}
(H \psi)_n = -\frac{1}{2(\delta x)^2}(\psi_{n+1}-2\psi_n+\psi_{n-1})&+V(x_n)\psi_n \nonumber \\ 
+\frac{i A(t)}{2 c~ \delta x}(\psi_{n+1}-\psi_{n-1})\,.
\end{align}


In both gauges the discretization of the Hamiltonian gives rise to a tridiagonal matrix. The length gauge produces a real symmetric-tridiagonal matrix and the velocity-gauge results in a Hermitian-tridiagonal matrix. The boundary conditions set the wavefunction to zero outside the computational region.  Consequently it is important to ensure that this region is sufficiently large to extract the probabilities of excitation and ionization without any spurious reflections that would compromise the numerical results. 


We also examined the central finite difference (CFD) discretization of the grid using 5, 7 and 9 point formulas. An $l$ point CFD discretization is ($\delta x^{l-1}$) accurate for the derivatives with respect to $x$. Matrix vector multiplication operations, which are the most time consuming part of several time-propagation schemes, could scale better by using higher orders of CFD. As an example, an $n \times n$ tridiagonal Hamiltonian matrix needs 3$n$ operations for matrix-vector multiplications while an $m \times m$ 9 point CFD matrix requires 9$m$ operations. However, for a given accuracy, the number of points $m$ could be much smaller than $n$, tipping the balance in favor of the higher order formula.  In the end, its the total number of floating point operations that determine the solution time. In addition, there is an advantage in using higher order CFD's in combination with coarser grids in that that the spectral range of the resulting matrices, i.e. the difference between highest and lowest eigenvalues of the matrix, could be significantly reduced.     
The impact of the spectral range on explicit propagation methods will be discussed in subsection~\ref{subsec:RSG}.

Finally, it is worthwhile pointing out that the Numerov method, which also produces tridiagonal matrices, is two orders of magnitude more accurate than the three point finite difference method and has been used with the lowest order Crank-Nicolson method for time propagation~\cite{kopal_numerical_1961,h.g._muller_efficient_1999,moyer_numerov_2004,esry_comparing_2018}. 

\subsection{\label{sec:CompParams} Computational Implementation}
In this study we began by examining the convergence of the probabilities of excitation to a few of the lower lying bound states. It is sufficient in this case to employ a box of $\approx$ 200 a.u. on each side of the origin. However, to obtain converged results for the continuum states, a box of  $\approx$ 800 a.u. on either side of the origin is necessary. 

A spatial step of $\delta x=0.1$ was employed, resulting in matrix sizes of $n=4001$ to $n=16001$. Laser interaction times varied for the tests but on average were about 1200 a.u.. The time-steps were varied from 1 to 0.001 a.u., according to the propagation method used. We chose a laser amplitude of $E_0$=0.1 and an angular frequency $\omega$=0.148. The value $\omega$=0.148 corresponds to five photon ionization. This is somewhat arbitrary, but is chosen here to show correspondence with the results obtained in \cite{j._javanainen_numerical_1988}.

These choices of parameters enabled us to perform all of the computational experiments on a desktop PC with Intel 3.4GHz Xeon(R) CPU's\footnote[2]{Certain commercial equipment, instruments, or materials are identified in this paper in order to specify the experimental procedure adequately. Such identification is not intended to imply recommendation or endorsement by the National Institute of Standards and Technology, nor is it intended to imply that the materials or equipment identified are necessarily the best available for the purpose.} in a practical amount of time.  The code was written in FORTRAN and compiled using the Intel-Fortran compiler with -Ofast optimization  The Intel MKL libraries were used in scalar mode to deliberately avoid any questions of the capabilities of the individual methods to employ OpenMP parallelization.  

We should emphasize that due to the variety and performance capabilities of platforms, compilers and libraries, the timings presented in this paper should be only regarded as an indication of the relative performance of the methods. 

\section{\label{sec:compmethods}Computational methods}
\subsection{\label{sec:cn}Crank-Nicolson}
The Crank-Nicolson (CN) method~\cite{j._crank_practical_1947} is an	implicit propagation/diffusion numerical method as the calculation of the solution at N$^{th}$ time-step  requires the solution of a set of linear algebraic equations.  Employing the second-order accurate version of the CN approximation in time, yields,
\begin{equation}\label{eq:CN}
 e^{-i \boldsymbol{H}(x,t) \delta t}= 
 \frac{e^{-i \boldsymbol{H}(x,t) \delta t/2}}
 {e^{i \boldsymbol{H}(x,t) \delta t/2}}
 \approx  \frac{[ 1 - i  \delta t/2 \boldsymbol{H}(x,t) ]}{[ 1 + i \delta t/2 \boldsymbol{H}(x,t) ]}\,,
\end{equation}
\begin{equation}\label{eq:CNSplit}
[ 1+ i \dfrac{\delta t}{2} \boldsymbol{H}(x,t)] \psi(x,t+\delta t)= [ 1 - i \dfrac{ \delta t}{2} \boldsymbol{H}(x,t) ]\psi(x,t).
\end{equation}

If we insert either of the finite-difference formulas of Eqs.\eqref{eq:tridiaglength} or \eqref{eq:tridiagvelocity} into Eq.\eqref{eq:CNSplit} , we obtain a tridiagonal set of linear equations.  These equations may be solved by a method which scales linearly with the number of unknowns. 

In higher dimensions, the coupling destroys the tridiagonal nature of the one-dimensional CN method, but it is still possible to derive second-order methods which scale reasonably well with matrix size \cite{kulander_time-dependent_1982}. We also note, that there are a number of methods which can solve a tridiagonal system faster than O(n) in parallel.  The reader is referred to Refs.~\cite{wang_parallel_1981,amodio_parallel_1992,mattor_algorithm_1995} for more details.  
\subsection{\label{sec:so}Split Operator}
Since the Hamiltonian can be split into a sum of time-independent and time-dependent parts, it is natural to consider propagation methods employing operator splitting to simplify the numerics. In all these  methods one is neglecting the commutator of operators that do not exactly commute. The non-commuting parts are proportional to some  power of the time step.  Thus, given small enough times steps, it is always possible to write the exponential as a simple exponential product, one for each operator in the Hamiltonian.  In practice, the size of the time step limits the accuracy and various $n^{th}$ order approximations are employed to make the approach numerically tractable, efficient and still sufficiently accurate for the time step chosen.

A well known, second-order accurate split operator (SO) method for the time evolution of the wavefunction is \cite{feit_solution_1982,de_raedt_product_1987,hermann_split-operator_1988,suzuki_fractal_1990,hochbruck_krylov_1997,Schneider2011,gonoskov_single-step_2016,jiang_efficient_2017}
\begin{equation} \label{eq:SO}
\begin{split}
\psi(x,t+\delta t)=& \exp[\frac{i \delta t \boldsymbol{V}(x,t)}{2}] \exp[-i \delta t \boldsymbol{H}_0(x)]\\ &\exp[\dfrac{i \delta t \boldsymbol{V}(x,t)}{2}]\psi(x,t).
\end{split}
\end{equation} \smallskip

Higher-order approximations could be derived by the method of fractal decomposition~\cite{suzuki_fractal_1990}.
 The 4th-order split operator, for example, is the more complicated operator \cite{hatano_finding_2005} \footnote[3]{Hatano and Suzuki~\cite{hatano_finding_2005}  have made a typo error in their equation (63); the factor of the middle term's exponentials remains ($1-4S_2$), not $S_2$. }
\begin{equation} \label{eq:SO4}
\begin{split}
&\hat{\mathbf{U}}(t+\delta t;t)=\\
&e^{\frac{i }{2} S \delta t \boldsymbol{V}(x,t+\frac{2-S}{2} \delta t)} e^{-i S \delta t \boldsymbol{H}_0(x)} e^{\frac{i }{2} S \delta t \boldsymbol{V}(x,t+\frac{2-S}{2} \delta t)}\\
&e^{\frac{i }{2} S \delta t \boldsymbol{V}(x,t+\frac{2-3S}{2} \delta t)} e^{-i S \delta t \boldsymbol{H}_0(x)} e^{\frac{i }{2} S \delta t \boldsymbol{V}(x,t+\frac{2-3S}{2} \delta t)}\\
&e^{\frac{i }{2} (1-4S) \delta t \boldsymbol{V}(x,t+\frac{1}{2} \delta t)} e^{-i (1-4S) \delta t \boldsymbol{H}_0(x)} e^{\frac{i }{2} (1-4S) \delta t \boldsymbol{V}(x,t+\frac{1}{2} \delta t)}\\
&e^{\frac{i }{2} S \delta t \boldsymbol{V}(x,t+\frac{3S}{2} \delta t)} e^{-i S \delta t \boldsymbol{H}_0(x)} e^{\frac{i }{2} S \delta t \boldsymbol{V}(x,t+\frac{3S}{2} \delta t)}\\
&e^{\frac{i }{2} S \delta t \boldsymbol{V}(x,t+\frac{S}{2} \delta t)} e^{-i S \delta t \boldsymbol{H}_0(x)} e^{\frac{i }{2} S \delta t \boldsymbol{V}(x,t+\frac{S}{2} \delta t)}\,,
\end{split}
\end{equation} \smallskip
where \(S=\frac{1}{4 - \sqrt[3]{4}}\). However, while the 4th-order splitting affords larger time-steps to be taken, the number of operations has increased five-fold as compared to the 2nd-order splitting. The trade-off between larger time-steps and larger number of operations did not lead to a performance advantage in our model problem.

A significant advantage of the SO approach is that it is an explicit time propagation method and does not require the solution of a set of linear algebraic equations to find the wavefunction at the next step. As is the case with CN, SO is more efficient in the length gauge, given that the matrix $V(x,t)$ is a local, diagonal matrix in configuration space. The remaining challenge is how to treat the exponential operator involving $H_0$. This is the computational bottleneck because a full diagonalization of $H_0$ can be quite expensive if the matrix is large. Even though our model enables the eigenvalues to be found quite efficiently, the eigenvectors are another matter and they are required to represent the exponential.  The eigenvector matrix is not only dense but is, in principle,  required for all the eigenvalues. 

A viable alternative is to treat the exponentiation of $H_0$ with either CN, fast Fourier transformation (FFT) or Lanczos iteration. The CN method leads to a set of tri-diagonal linear equations just as before. In the FFT approach, $H_0$ is split into its kinetic energy and potential energy parts and FFT is used to treat the kinetic energy essentially exactly.  For the current problem we saw no computational advantage in using the FFT over CN.  Either the direct use of CN or Lanczos is computationally easier and potentially more accurate.
We describe the Lanczos method in some detail in the next section.

We mention a third approach, called the real space split operator method \cite{de_raedt_product_1987}, where splitting the second derivative in real space has been used to bypass the use of the fast Fourier transform \cite{press_numerical_2007} of the kinetic energy. This does not destroy the second order accuracy of the method.  Since the Hamiltonian in question is tridiagonal, the splitting leads to an expression where the exponential can be written as a sum of overlapping ``even" and ``odd" $2\times 2$ matrices. 
\begin{align*}
\begin{bmatrix}
    a_1 & b_1 &  &  &  \\
   b_1 & a_2 & b_2 &  &  \\
    	 & b_2 & a_3 & b_3  &  \\
     & & b_{3} & a_{4} & b_{5}\\
     &  &  & b_{5}  & a_{5} \\  
\end{bmatrix} =
&\begin{bmatrix}
    a_1 & b_1 &  &  &  \\
   b_1 & a_2/2 &  &  &  \\
    	 &  & a_3/2 & b_3  &  \\
     & & b_{3} & a_{4}/2 & 0\\
     &  &  & 0  & 0\\
\end{bmatrix}+\\
&\begin{bmatrix}
    0 & 0 &  &  &  \\
   0 & a_2/2 & b_2 &  &  \\
    	 & b_2 & a_3/2 &   &  \\
     & & & a_{4}/2 & b_{5}\\
     &  &  & b_{5}  & a_{5} \\  
\end{bmatrix}\,.
\end{align*}
If we now insert this into the exponential and employ a $2^{nd}$ order accurate splitting of the exponential we obtain,
\begin{equation} \label{eq:SO2}
\begin{split}
\exp[-i\delta t H_0]=& \exp[\frac{i\delta t}{2} \boldsymbol{H}_0^ {\rm{odd}}] \exp[-i\delta t \boldsymbol{H}_0^{\rm {even}}]\\
& \exp[\frac{i\delta t}{2} \boldsymbol{H}_0^{\rm {odd}}],
\end{split}
\end{equation}
Since $H_0^{\rm ``even"}$ and $H_0^{\rm ``odd"}$ each consist of non-overlapping $2\times2$ matrices, it is easy to diagonalize them and to then propagate the solution using operations that only require the successive action of a set of $2\times2$ matrices on a vector.  By employing this even-odd splitting approach we are neglecting commutators between different parts of the  discretized kinetic energy operator. We have found that this approach only becomes reasonable for quite small time-steps compared to other techniques.

A time comparison for the various splitting methods will be shown in section \ref{sec:data}.

\subsection{\label{sec:lncz}Lanczos Iteration}
The Lanczos Iteration method was initially developed to find the smallest and largest eigenpairs of a large, sparse, $n \times n$ symmetric matrix~\cite{cornelius_lanczos_iteration_1950,paige_computation_1971,cullum_lanczos_1985}. Lanczos-type solvers have also been used to deal with non-symmetric matrices (see, for example, \cite{axelsson_iterative_1994} and references therein). It has been demonstrated that these eigenvalues converge in far fewer than $n$ steps. A significant advantage of the method is that the major computational step involves the multiplication of the matrix on a known vector plus a few scalar products.  For a sparse matrix, this can be done in $O(M)$ multiplications, where $M$ is the number of non-zero matrix elements.  

In effect, the Lanczos method may be viewed as reducing the large, $\rm{\mathbf{H}}_{n\times n}$ matrix to a smaller, $\rm{\mathbf{H}^{(\Lambda)}_{m\times m}}$ tridiagonal matrix, where ideally $m \ll n$ for the eigenpairs of interest. If the iteration continues until $m=n$, the eigenvalues and eigenvectors of $\rm{\mathbf{H}^{(\Lambda)}}$ would be identical to those of the $\rm{\mathbf{H}}_{n\times n}$. A naive implementation of the Lanczos iteration can lead to linear dependence which has a number of undesired side effects~ \cite{cullum_lanczos_1985,axelsson_iterative_1994}.  To circumvent linear dependence often requires additional and expensive mathematical operations.  

The transformation between the two representations is given by:
\begin{equation}\label{eq:TQH}
\mathbf{H}^{(\Lambda)}=\mathbf{Q}^T\mathbf{H}\mathbf{Q}\,,
\end{equation}
where $\rm{\mathbf{Q}}_{n \times m}$=[$\ket{q_1}~\ket{q_2}~...~\ket{q_m}$] are the Lanczos vectors at the $m^{th}$ step of the process.  One can view these vectors as linear combinations of the so-called Krylov subspace vectors-
\begin{align}\label{eq:krylovsubspace}
K(H,q,m)&=\rm{span}\{\ket{q_1},\textbf{H}\ket{q_1}, \textbf{H}^2 \ket{q_2},..., \textbf{H}^{m-1}\ket{q_1}\} \nonumber \\
&=\rm{span}\{\ket{q_1}, \ket{q_2}, \ket{q_3},..., \ket{q_m}\},
\end{align}
where,
\begin{equation}
\label{eq:lanczos_vectors}
\beta_{k+1} \ket{q_{k+1}}=(\mathbf{H}-\alpha_k \mathbf{I})\ket{q_k}-\beta_{k-1} \ket{q_{k-1}}. 
\end{equation}
The vectors in~Eq.\eqref{eq:lanczos_vectors} form an orthonormal set.  The recursion relation may be started with any vector and continued until the desired eigenvalues are found to sufficient accuracy.

The application of the Lanczos method to time propagation is not directly related to the question of determining the eigenvalues \cite{tae_jun_park_unitary_1986,leforestier_comparison_1991}. The process may be stated as follows; let the first Lanczos vector, $\ket{q_1} = \ket{\psi{(x,t)}}$.  How can we determine a small set of additional vectors, $\ket{q_2} \dots \ket{q_m}$, which effectively span the new subspace defined by $\ket{\psi{(x,t+\delta t)}}$ and provide a representation of the exponential function over the time-step? We assume the time-step is small enough that we can approximate the interaction of the electrons with the field by using the field's value  at the midpoint of the time-step. Then the time evolution operator, $\hat{\textbf{U}}$, may be approximated as
\begin{equation}\label{eq:timeevolutionop}
\hat{\textbf{U}}(t+\delta t|t) = \exp{[-i \hat{\textbf{H}}(x,t+\frac{\delta t}{2}) \delta t]}.
\end{equation}

The size of the the Krylov subspace matrix, \textbf{Q}, is determined by the sufficiently accurate approximation of \textbf{H} by the tridiagonal matrix
\begin{equation}\label{eq:KrylovMatrix}
\textbf{H}^{(\Lambda)}=
\begin{bmatrix}
    \alpha_1 & \beta_2 &  &  &  \\
   \beta_2 & \alpha_2 &  &  &  \\
    	 &  & \ddots &   &  \\
     & &  & \alpha_{m-1} & \beta_{m}\\
     &  &  & \beta_{m}  & \alpha_{m} \\  
\end{bmatrix}.
\end{equation}

The most computationally demanding step in the process involves the multiplication of \textbf{H} onto the previously computed Lanczos vector. In our case of the 1-D propagation, this comes down to multiplying a tridiagonal matrix by a vector which is an $O(3n)$ process.

To compute the wavefunction at the next time using the Lanczos iterations we replace the time evolution operator of \textbf{H} by its approximation in the Krylov subspace,
\begin{equation}\label{eq:lanctimeevol}
\hat{\textbf{U}}^{L}=\exp(-i ~\hat{\textbf{H}}^{(\Lambda)} \delta t).
\end{equation}
Therefore, $\hat{\textbf{U}}^L$ is restricted to the (hopefully) smaller dimensional Krylov subspace and is evaluated by direct diagonalization of the tridiagonal matrix produced by transforming \textbf{H} to the Lanczos basis,
\begin{equation}
\hat{\textbf{U}}^L= \sum_i \ket{\lambda_i}\exp(-i \lambda_i \delta t)\bra{\lambda_i}\,, 
\end{equation}
where $\ket{\lambda_i}$ denotes the eigenvector of \textbf{H}$ ^{(\Lambda)}$ with eigenvalue $\lambda_i$.
The propagated wave is then
\begin{equation} \label{eq:lncz}
\ket{\Psi(x,t+\delta t)}= \sum_i \ket{\lambda_i} \exp(-i\lambda_i\delta t)\braket{\lambda_i}{\Psi(x,t)}\,.
\end{equation}
In matrix notation this is equivalent to 
\begin{equation}
\bm{\Psi}( t+\delta t ) = \textbf{Q} \mathbf{\Lambda}^T\exp(-i \delta t~ \text{diag} [\lambda_1~...~\lambda_m]) \mathbf{\Lambda} \textbf{Q}^T \bm{\Psi}(t)\,,
\end{equation}
where \(\mathbf{\Lambda}=[~\ket{\lambda_1}~ \ket{\lambda_2}~...~\ket{\lambda_m}~]\).  
\\
\subsubsection{Algorithm}
\begin{table}[ht]
\begin{center}
\begin{tabular}{ |l| } 
 \hline
 \textbf{Algorithm for Lanczos Propagation}     \\
 \hline
 \textbf{initialize}:
 $\ket{q_{0}} = 0;~\beta_1=0;~k=1$ \\
 $n_{\text{max}}=\text{Lanczos maximum vector number/iteration}$ \\
 $\epsilon=\text{error threshold}$ \\
 $N  = \sqrt{ \braket{ \Psi(x,t)}{\Psi(x,t)} }$\\
 $~\ket{q_1} = \ket{\Psi(x,t)} / N$ \\
 \textbf{do while} \textit{k} $< n_{\text{max}} $ \\ 
 ~$\alpha_k$ = $\bra{q_k}\mathbf{H}\ket{q_k}$  \\
 ~ Calculate eigenvalues ($\lambda_i$) and eigenvectors ($\ket{\lambda_i}$)\\ ~ of tridiagonal matrix \textbf{H}$^{(\Lambda)}$ (i $\in \{1,\dotsc,k\}$); \\
 ~$\ket{\Psi'_{k}}=\sum_{ij} \ket{\lambda_{i}}~\exp(-i \lambda_{i} \delta t) 	~\braket{\lambda_{i}}{q_j}\braket{q_j}{\Psi(x,t)} $ \\
~ \textbf{if} $( ||~\ket{\Psi'_{k}} -\ket{\Psi'_{k-1}}~||_2< \epsilon$)~ \\
~~  $\textbf{converged} = \textnormal{True}$ \\
~~	\textbf{break}\\
~ $\ket{r_k} = (\mathbf{H}-\alpha_k \mathbf{I})\ket{q_k} - \beta_{k-1} \ket{q_{k-1}}$ \\
~ $\beta_{k+1}=\sqrt{\braket{r_k}{r_k}}$ \\
~ $\ket{q_{k+1}} = \ket{r_k}/ \beta_{k+1}$ \\
~ $k=k+1$ \\
\textbf{end 1st while loop}\\
\textbf{while .not. converged}\\
~$\delta t=\delta t /2$\\
~recalculate $\Psi'_{k}$ and $\Psi'_{k-1}$ \\
\textbf{end 2nd while loop}\\
$\psi(x,t+\delta t) = \sum_j\braket{q_j}{\Psi'_k}$ \\
 \hline
\end{tabular}
\caption{}\label{table:lanczoslgorithm}
\end{center}
\end{table}
In our application of the Lanczos method, we adjusted the size of the time-step during the propagation to yield accurate results using the smallest number Krylov space vectors.  For simplicity, this variable time-step approach is still referred to as Lanczos propagation in this paper. A brief description is as follows; we start the process by setting
\begin{align}
\label{eq:initial_vector}
&\ket{q_1} = N^{-1} \ket{\Psi(x,t)}\,, \\
&N = \sqrt{\braket{\Psi(x,t)}{\Psi(x,t)}}\,.\nonumber
\end{align}
This is a normalized version of the wavefunction from the previous time-step.
The Lanczos vectors are generated from the three-term recursion relationship,
\begin{align}\label{eq:lancfundemental}
\ket{r_k}=\beta_{k+1} \ket{q_{k+1}} = ( \mathbf{H} -\alpha_k ) \ket{q_{k}}
- \beta_{k-1} \ket{q_{k-1}}\,.  
\end{align}
Due to the orthonormality of the $\ket{q_k}$'s,
\begin{align}
\label{eq:alpha}
\alpha_k = \bra{q_k}\mathbf{H}\ket{q_k}.
\end{align}
Having found $\alpha_k$, from Eq.(\ref{eq:alpha}) we can find $\beta_{k}$ from
\begin{align}
\beta_{k+1} &=\sqrt{\braket{r_k}{r_k}} \nonumber \\
\ket{q_{k+1}} &= \frac{\ket{r_k}} 
{\beta_{k+1}}\,.
\end{align}

In setting the maximum number of Lanczos vectors to a modest value, for example, 20 vectors, the iteration process will or will not converge for the selected time-step.  If it converges, we continue to the next step.  If it does not converge, the time-step is halved and the convergence is again tested.  This process is repeated until the propagated vector is sufficiently accurate to continue. 
Note that one does not need to compute any new Lanczos vectors for the reduced time-step.
All that needs to be done is to re-evaluate the exponential on the vector, a numerically cheap operation.
Pseudocode of the algorithm we implemented is given in Table~\ref{table:lanczoslgorithm}.

As with all applications of the Lanczos method, one needs to pay attention to the possible loss of orthonormality during the iteration process.  This loss of linear independence can lead to disastrous results. We found that if the initial time-step is taken too small, the first and second Lanczos vectors are not sufficiently linearly independent to produce stable results.  By re-orthogonalizing those two vectors, the rest of the process proceeds smoothly, without any need for additional Gram-Schmidt steps.

The algorithm as described here is robust and self-correcting. However, a slightly faster algorithm could be written for a fixed number of Lanczos vectors associated with a time scale, $\delta t$, known in advance to converge.  This would not only eliminate the need to check convergence at every step but would also reduce the need to tridiagonalize the Lanczos matrix multiple times. Such a scheme would necessarily need to be tested for convergence by an a priori adjustment of $\delta t$; the same statement can be made about the operator splitting and Crank-Nicolson methods. 

\subsubsection{Split-Operator+Lanczos}
It is possible to combine the splitting techniques of subsection \ref{sec:so} with the time-adjusting Lanczos scheme described above. The Hamiltonian in the length-gauge in our 1-D model problem lends itself to such a splitting.

By splitting the time evolution operator in the length-gauge into time independent and time-dependent exponentials, one produces, in second order, an expression consisting of a single, time independent, tridiagonal matrix sandwiched between two diagonal time-dependent exponentials.  The time independent, tridiagonal matrix is treated via the Lanczos propagator and its diagonal scaling used to compute the required vector. We call this the Split-Operator+Lanczos method. 

Even with this tact, the number of Lanczos vectors needed for convergence at a given time-step may change due to the diagonal, time-dependent scaling.  The consequence is still a time-adjusting scheme. The time-step at each time iteration, for a fixed maximum number of Lanczos vectors, has to be adjusted to ensure convergence.  As mentioned before, one could employ a fixed number of Lanczos vectors for all time-steps, if that could be determined prior to the propagation.  Finally, higher order splittings could be used to allow larger time-steps, but the number of operations will also increase, which could result in overall longer run-times for the same accuracy.       
 
\subsection{\label{sec:cheby}Chebychev propagator}
The final method we examined was the so-called Chebychev propagator \cite{talezer_accurate_1984,kosloff_time-dependent_1988}. This approach has been used in the chemical physics community to treat a number of problems but, to our knowledge, has not seen widespread use in the atomic and molecular physics literature. The basic idea is to expand the short time evolution operator \(\boldsymbol{\hat{U}} = \exp(-i \boldsymbol{H} dt )\) in terms of a complex version of the Chebychev polynomials of the first kind.  
Since the Chebychev polynomials are defined on the interval \(x \in [-1,1],\) this requires the Hamiltonian to have eigenvalues inside the unit circle.  This can be accomplished if we can estimate the spectral range, $\Delta=E_{\text{max}}-E_{\text{min}}$ of $ \boldsymbol{H}$, where $E_{\text{max}}$ and $E_{\text{min}}$ are respectively, the largest and smallest eigenvalues of $ \boldsymbol{H}$.  Then using,
\begin{equation}
\hat{\boldsymbol{H}}_{\textnormal{norm}} = 2 \frac{\hat{\boldsymbol{H}}-E_{\textnormal{min}} \boldsymbol{I}}{\Delta}-\boldsymbol{I},
\end{equation}
the expansion coefficients can be shown \cite{talezer_accurate_1984} to be
\begin{equation}\label{eq:coefCheb}
a_k=(2-\delta_{0k})\exp[(-i \frac{\Delta}{2}+E_{\textnormal{min}} )dt] J_k(\frac{\Delta}{2} dt)\,.
\end{equation}
where $J_k$ are the Bessel functions of the first kind.   
If the spectrum of the Hamiltonian does not vary greatly in time, as is the case for the 1-D problem we discuss here, then a single set of coefficients $a_k$ can be obtained and truncated to a desired threshold limit for all time-steps. The propagation is performed in this case with the $\hat{\boldsymbol{H}}_{\text{norm}}$ and a normalized $\psi_0$, using the recursive relationship between Chebychev polynomials. Below is a quick outline for the Chebychev propagator which follows the more detailed explanation given in Goerz \cite{goerz_optimizing_2015}. 
\begin{table}[ht]
\begin{center}
\begin{tabular}{ |l| } 
 \hline
 \textbf{Algorithm for Chebychev Propagation}     \\
 \hline
 \textbf{initialize}:
 $\Delta=\textnormal{spectral range of }\hat{H}$ \\
 $E_{\textnormal{min}}=\textnormal{minimum eigenvalue of }\hat{H}$ \\
$\beta=\frac{\Delta}{2}+E_{\textnormal{min}}$ \\
$[a_0 ...a_n]=\textnormal{Coefficients (see Eq.} \ref{eq:coefCheb}$) \\
 $d=\frac{\Delta}{2};~\beta=d+E_{\textnormal{min}} $ \\
 $\ket{v_0}=\ket{\psi(x,t=t_0)} $ \\
 $ \ket{\omega_0} = a_0 \ket{v_0} $\\
 $\ket{v_1} = -\frac{i}{d}(\hat{\mathbf{H}} \ket{v_0}-\beta \ket{v_0})$\\
 $\ket{\omega_1} = \ket{\omega_0}-a_1 \ket{v_0}$\\
 \textbf{for} \textit{k} = 2 : n  \\ 
 ~$\ket{v_k}$ = $-\frac{i}{d}(\hat{\mathbf{H}} \ket{v_{k-1}}-\beta \ket{v_{k-1}})+\ket{v_{k-2}}$  \\
 ~$\ket{\omega_k} = \ket{\omega_{k-1}}-a_k \ket{v_{k}}$  \\
\textbf{end for}\\
\textbf{return} $\ket{\psi(x,t+dt)} = \ket{\omega_n}$\\
 \hline
\end{tabular}
\caption{}\label{table:chebyalgorithm}
\end{center}
\end{table}

In Table \ref{table:chebyalgorithm} we present pseudocode for the propagation using Chebychev expansion. Here it is assumed that the coefficients of Eq.\eqref{eq:coefCheb} are already calculated for the time-step. 

In the next sections, we analyze the performance and results of the method comparisons.

\section{\label{sec:data}Results}
\subsection{\label{subsec:results}Performance}

\begin{figure}[ht]
\includegraphics[scale=0.36]{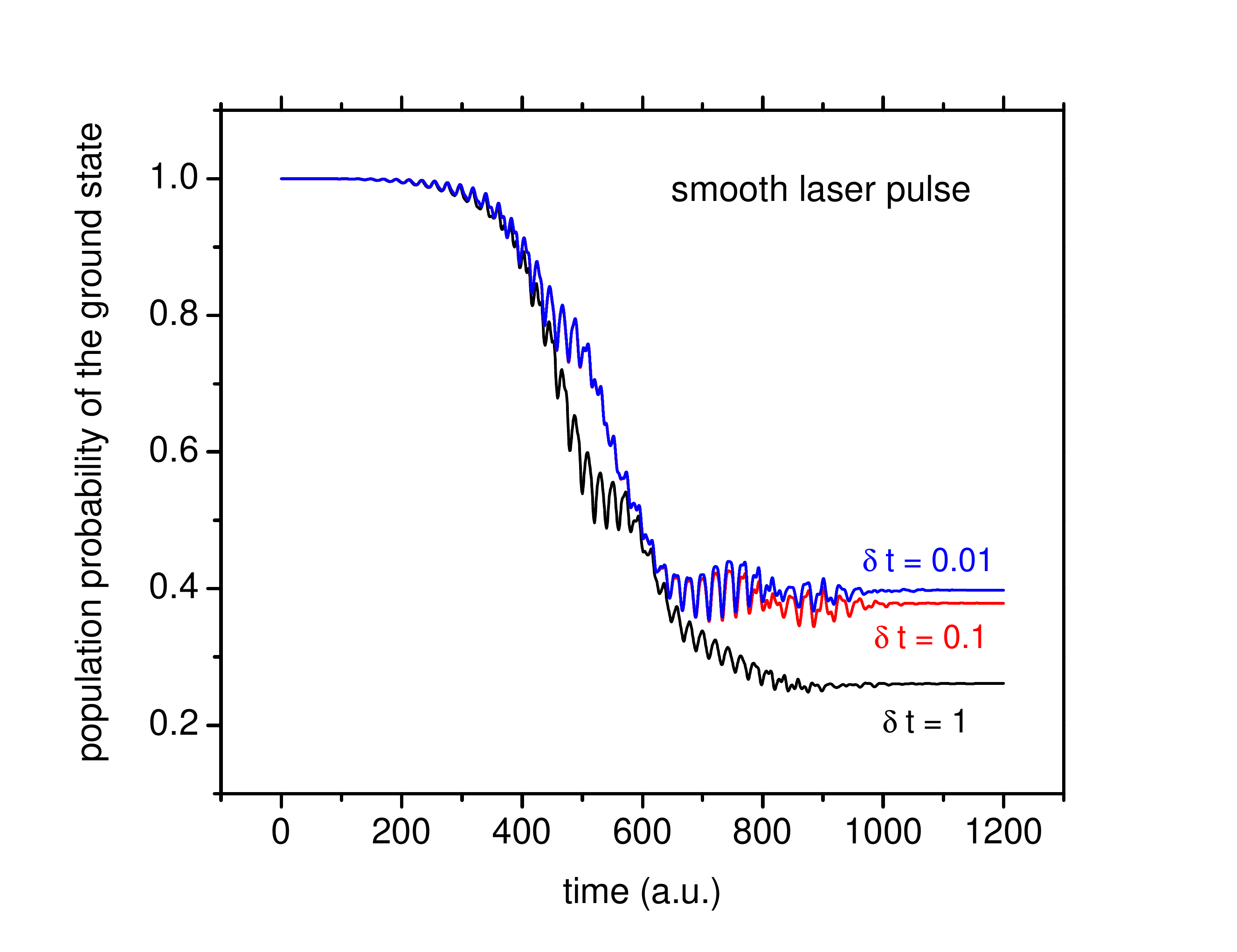} 
\caption{Convergence of the population of the ground state over time using the Crank-Nicolson method, in the length gauge, for a smooth laser pulse with $E_0$=0.1, $\omega$=0.148, $T$=1200. There are $2000$ points on either side of $x=0$.}
\label{fig:populationgroundconverge}
\end{figure}

\begin{figure}[ht]
 \centering
 \includegraphics[scale=0.36]{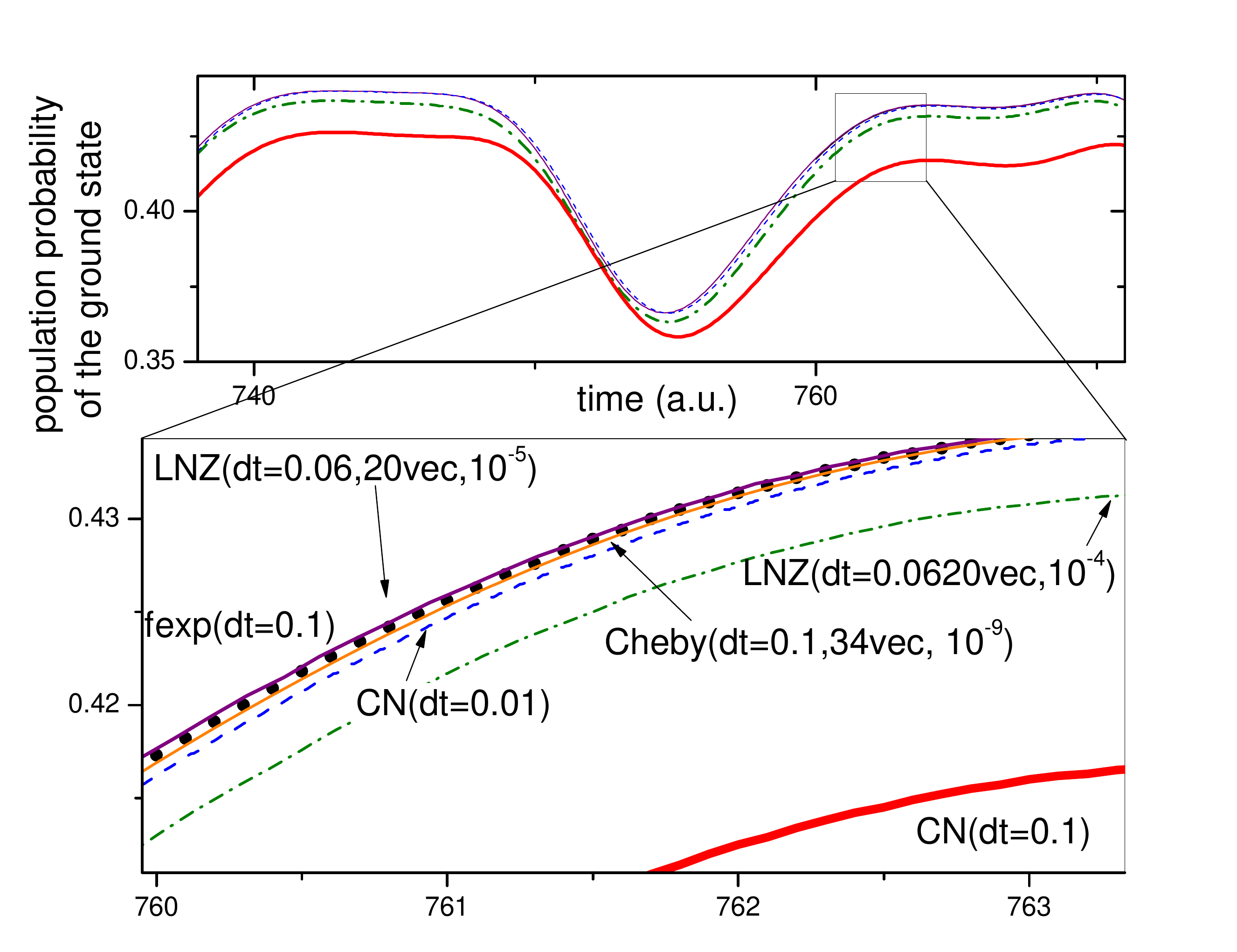} \caption{A magnified view of a section of Fig.~\ref{fig:populationgroundconverge}. The region encompassed by the small square in the upper figure is enlarged in the lower figure. Here CN refers to Crank-Nicolson method, fexp refers to full diagonalization of the exponential at each time step, and LNZ refers to Lanczos propagation.
 }
 \label{figure:CloseUp}
\end{figure}

In this section we examine the numerical performance, accuracy and timing of the propagation methods mentioned in section \ref{sec:compmethods}. Low-order methods can often be quite efficient for a given time-step but might require very small time-steps to achieve high accuracy. Note also that performance and accuracy can depend on both the spatial and temporal discretization and the balance between the two.  Consequently, it is important to examine both parameters before drawing any conclusions.  

In the present context, the term accuracy is measured by the convergence of the populations of the bound and continuum states as a function of time for some fixed spatial and time-step.  

\subsubsection{Excitation Results for a Spatial Grid $x=0.1$}
For the ground and low-lying bound states examined in section, it is sufficient to use a computational region that is $200$ a.u. to either side of the origin and a spatial grid of $0.1$ a.u.  This spatial grid is sufficient to ensure that the populations of the ground and low lying excited states would not change at the end of the pulse if the grid was refined further. This grid results in a $4001 \times 4001$ tridiagonal matrix. As a baseline, we first examined the evolution of the ground state population, $|\braket{\psi}{\psi_0}|^2$, in time using the CN method for a smooth pulse in the length gauge.

In Fig.~\ref{fig:populationgroundconverge} the CN results are shown for three different time-step sizes.
The time-steps in this graph increase top to bottom and are written next to each curve. Any line with  $\delta t< 0.01$ will lie on top of the upper most curve (blue curve online). We are using a solid black line for $\delta$t=1, a solid thick (red online) for $\delta$t=0.1 and a dashed thin line (blue online) for $\delta$t=0.01. The figure indicates that for the CN method converged results for the probabilities require a $\delta t=0.01$ or smaller.

Physically, the oscillations in Fig.~\ref{fig:populationgroundconverge} are a consequence of the change in the population of the ground state of the system as the intensity of the laser field oscillates in time over a period. In the absence of ionization the overall population of the ground state will decrease and the excited states increase over time until the populations equilibrate. 

The upper panel of Fig.~\ref{figure:CloseUp} displays a magnified portion of Fig.~\ref{fig:populationgroundconverge} while the lower panel displays two new plots using the Lanczos and Chebychev methods.  The Lanczos calculations used a maximum of 20 vectors (i.e., $\ket{v_k}$ in Eq.~\ref{eq:lancfundemental} for $k \in \{1,\dotsc,20\}$), with an initial $\delta t = 1.0$.  Two different convergence limits were used. The dash-dot line (green online) used a tolerance to 10$^{-4}$, while the solid line (purple online) used a tolerance of 10$^{-5}$. The adaptive time-step technique employed in the  Lanczos method reduced the  initial $\delta t= 1.0$ to $\delta t=0.0625$ to attain or exceed convergence with this number of Lanczos steps. For completeness, we also show a plot of the result of applying the fully diagonalized time-dependent Hamiltonian~Eq.~\ref{eq:timeevolutionop}, using a $\delta t=0.1$ (black scattered dots) and the result of the Chebychev propagator(thin solid line, orange online) using 34 vectors with tolerance of 10$^{-9}$ and time-step of $\delta t=0.1$. 
 
 In Table \ref{table:timingcompare}, we have listed the average propagation times for each of the methods, as performed serially on a desktop PC with CPU of 3.4 GHz and with an Intel-Fortran compiler with -Ofast optimization flag. The timings are for a spatial grid $\delta x =0.1$ a.u., in a box of size 4001 grid points, centered at zero. All other parameters are as in Fig.~\ref{fig:populationgroundconverge}.  The total time of propagation is 1200 a.u. For Lanczos iteration a maximum iteration number of 20 was used and the error threshold was set to 10$^{-5}$. Note, other settings can also effect the timing of Lanczos and Chebychev iterations, but again the purpose here is to achieve convergence in the probability of the populations of the ground and excited states.

\begin{table}[ht]
\begin{tabular}{|l|c|r|r|}
\hline
Method    & Time-Step  & Avg. propagation    \\ 
&$\delta t$&time (sec.) \\
\hline
Crank-Nicolson & 0.01 & 29 \\ 
\hline
Split-Operator$^{\text{2nd}}$ & 0.5 & 32 \\
\hline
Split-Operator$^{\text{4th}}$ & 1.0 & 100 \\
\hline
Even-Odd-Split-Operator & 0.001 & 180 \\
\hline
Lanczos   & 1.0~-~0.0625 & 14 \\
\hline
Split-Operator$^{\text{2nd}}$+Lanczos & 1.0~-~0.125 & 10 \\
\hline
Split-Operator$^{\text{4th}}$+Lanczos & 1.0~-~0.25 & 24 \\
\hline
Chebychev & 0.16 & 8 \\
\hline
\end{tabular}
 \caption{Average propagation times using different propagation methods. In all of the test cases the spatial grid was fixed at $\delta x =~0.1$, which results in a Hamiltonian matrix of size $N=4001$ having $2000$ points on either side of $x=0$. $\delta t$ was chosen so the probability of the ground state converged to a consistent profile as $\delta t\rightarrow 0$. }\label{table:timingcompare} 
\end{table}

\begin{figure}[ht]
 \centering
 \includegraphics[scale=0.36]{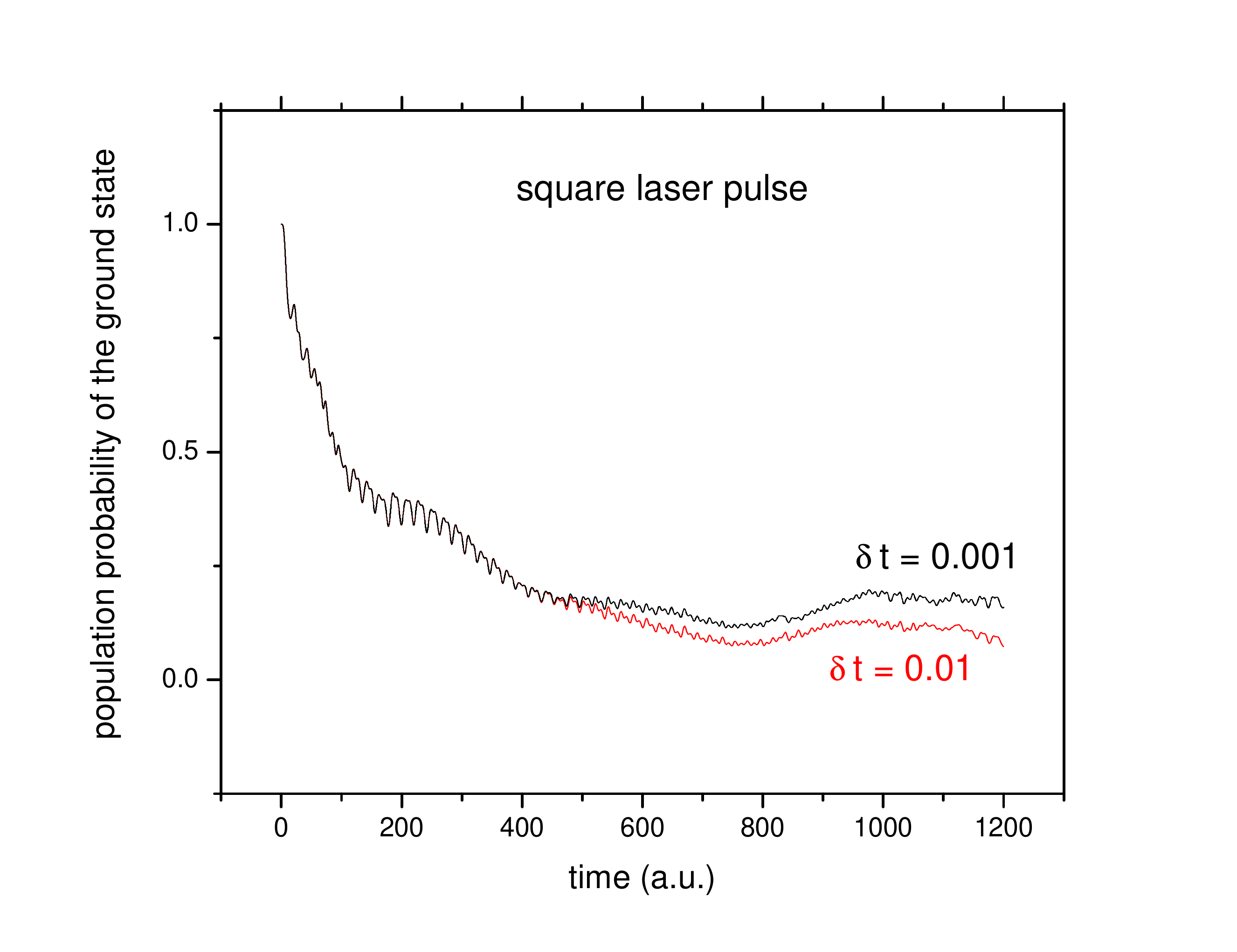} \caption{Same as Fig.~\ref{fig:populationgroundconverge} but using a square pulse. Here the time-step is shown for the Crank-Nicolson method.}	     
\label{figure:square_ground}
\end{figure}

We have also tested the influence of a square wave pulse~Eq.~(\ref{eq:E(t)_square}) on the convergence of the probabilities of low-lying states in time.  The square wave pulse shows an even greater dependence of the Crank-Nicolson needing small time-steps for high accuracy. Figure~\ref{figure:square_ground} shows that a square wave pulse drives down the population of the ground state even more rapidly than a smooth pulse and it is necessary to choose $\delta t=0.001$ in the CN method to achieve satisfactory convergence. On the other hand, the Lanczos method remains as accurate and as efficient as in the smooth pulse. 

\subsubsection{Reducing the Spatial Grid}
\label{subsec:RSG}
Although it is not necessary to reduce the spatial grid to below $\delta x=0.1$ to achieve converged results in time for this problem, we felt it was important to show data illustrating the performance of the methods with smaller spatial grid sizes. As the grid is refined for the tridiagonal matrix, the spectral radius \footnote{spectral radius of a matrix is the largest absolute value of its eigenvalues. } of the discretized Hamiltonian can become quite large. The size of the spectral radius influences many explicit time propagation schemes. There are ways to ameliorate these effects using preconditioning~\cite{axelsson_iterative_1994}. Combining higher order discretization methods with a coarser grid that preserves accuracy could also help in reduction of the spectral range \footnote{spectral range is the difference between lowest and highest eigenvalues of a matrix.}. As we discussed earlier in section \ref{subsec:discretizinggrid}, utilizing higher order finite difference discretization could boost performance.

Here we make a brief comparison between different spatial grid sizes made with $l$-point central finite difference (CFD) discretization methods, where $l$=3 or 9.
\begin{figure}[h]
 \centering
 \includegraphics[scale=0.36]{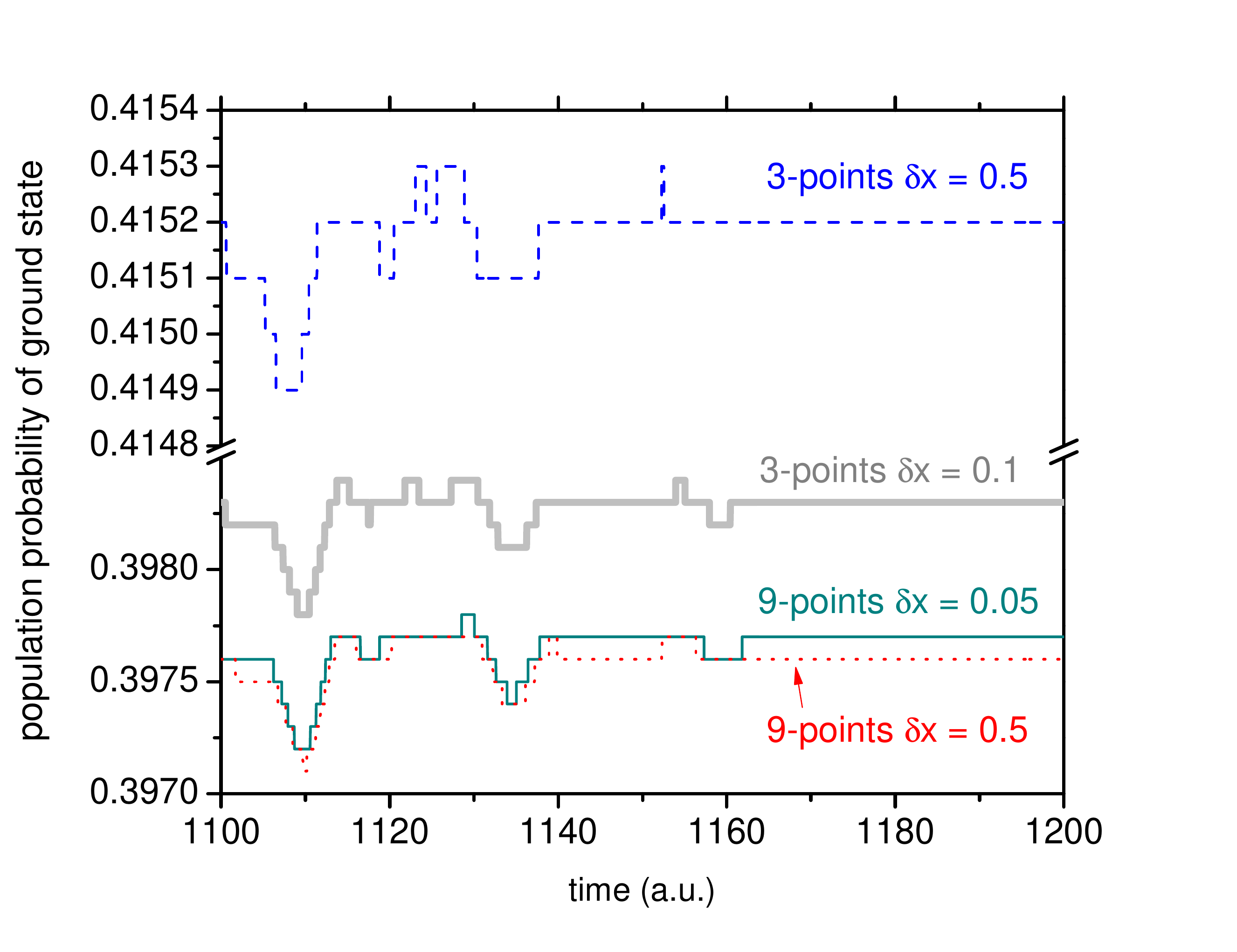} \caption{Comparison of a three and nine point finite difference method for different $\delta x$ while using fixed $\delta t$=0.01 a.u.}
 \label{figure:gridpoints}
\end{figure}
In Fig.\ref{figure:gridpoints} we show the convergence of the same probability as in previous figures for spatial grids made with higher order central finite difference (CFD) discretization scheme. Here only the last 100 seconds of smooth-pulse on is shown and all the curves have been made using spatial grids with $\delta x$ ranging from 0.5 to 0.05, but a fixed $\delta t$=0.01. The  thick solid gray curve is the result of time propagation with tridiagonal Hamiltonian and $\delta x$=0.1, the scheme that has been the focus of this paper. In order to better distinguish the 3 bottom curves, the bottom half of the plot is log scaled in probabilities. The ground state population probability is essentially converged to $\approx$0.001 by using either tridiagonal with $\delta x$=0.1 or a 9-point CFD's (nonu-diagonal) with $\delta x=$0.5. However, the size and the spectral range of matrices made with higher order CFD's could be smaller. This results in better performance of explicit propagation methods such as Lanczos and Chebychev.

We emphasize that both spatial and temporal grids are important in convergence of a TDSE problem.
However, in this paper our goal is mostly to compare methods of time propagation. Thus for the sake of simplicity we only restrict ourselves to mainly tridiagonal matrices. Additionally, we have focused primarily on the temporal aspects of the convergence criteria, given a fixed spatial grid accuracy. That said, we next examine the effects that the grid size and the matrix sizes have on the Split-Operator + Lanczos, which could be the fastest and most efficient method. 

In Table~\ref{table:timesfixedvectorsbiggerbox} we display what happens if $\delta x$ is reduced, using the tridiagonal matrices. Here we apply the split-operator+Lanczos approach, using an adaptive time-step with a fixed maximum number of vectors.  Notice that a smaller $\delta x$ requires a larger number of Lanczos iterations and a smaller $\delta t$ to achieve convergence.  This is a direct consequence of the larger spectral radius of the discretized Hamiltonian. The numbers displayed for $\delta t$ in column two are representative of the minimum time-step automatically chosen by the Lanczos propagation algorithm. At each time-step the method will automatically choose $\delta t$ to converge and it is impossible to predict in advance what the size of that step would be. The Lanczos method is more efficient than CN at the larger spatial steps and comparable to it at the smaller steps.

\begin{table}[ht]
\begin{center}
Timings for Split-Operator + Lanczos with fixed maximum number of Lanczos Iterations and an adaptive time-step
\end{center}
\begin{tabular}{|l|c|c|c|}
\hline
$\delta x /N$   &~~$\delta t$           &~~ Iterations        &~~ computation time \\
\hline
     &~0.0625			   &15					 & 11 sec.					    \\
 0.1/4001 &~0.125	   			   &20					 & 10 sec.   					\\
     &~0.25				   &50					 & 14 sec.	                     \\      
\hline  
&~0.03125	&15	&62 sec. \\
0.05/8001&~0.03125	&20	&63 sec. \\
&~0.125	&50	&85 sec.                      \\
\hline
\end{tabular}
\caption{Column one displays the size of the spatial grid, $\delta x$ and the matrix $N$. The initial time-step is $\delta t = 1.0$ but is automatically reduced (column two) according to the convergence criterion (see~Fig.~\ref{fig:populationgroundconverge}) and the maximum number of Lanczos iterations (column three) allowed in the computations. The propagated solution is converged to a tolerance = $10^{-5}$. The computation times illustrate the sensitivity to the spectral range of the matrices.}
\label{table:timesfixedvectorsbiggerbox}
\end{table}
 
In Table~\ref{table:timesfixeddtbiggerbox} we fix both the time-step and the maximum number of Lanczos iterations. Again, the Lanczos and CN methods are comparable at the larger spatial step but the spectral range of the Hamiltonian matrix for the smaller spatial step sizes slows down the Lanczos considerably if we do not employ an adaptive time-stepping scheme. The Chebychev propagator also suffers from dependence on the spectral range of the Hamiltonian for small step sizes.

\begin{table}[ht]
\begin{center}
Timings for Split-Operator + Lanczos with a fixed number of Lanczos iterations and a fixed time-step.
\end{center}
\begin{tabular}{|l|c|c|c|}
\hline
$\delta x/N$   &~~\space$\delta t$           &Iterations         &~~ computation time \\
\hline
& 1		& 106	& 43 sec. \\	& 0.5	    		   & 54					   & 24 sec. 		  \\
0.1/4001 	& 0.25				   & 30					   & 25 sec. 		   \\
		& 0.1				   & 15					   & 35.5 sec.			\\
		& 0.05				   & 10					   & 1.2 min.            \\ \hline     
		&~0.5                  & 380                   & 16 min.              \\ 
		&~0.25                 & 94                    & 2.8 min.              \\ 
0.05/8001	&~0.1                  & 38                    & 3.2 min.               \\ 
		&~0.05                 & 20                    & 5 min.                  \\ 
		&~0.025                & 13                    & 8.5 min.                 \\ \hline
\end{tabular}
\caption{First and second columns as in Table~\ref{table:timesfixedvectorsbiggerbox}. The third column shows the number Lanczos iterations required to converge the results to the same profile as in Fig.~\ref{fig:populationgroundconverge}. The last column displays the computation times.}
\label{table:timesfixeddtbiggerbox}
\end{table}

  One might wonder whether the slowing down of the Lanczos propagation with the decrease of spatial step size, $\delta x$, is intrinsically due to the need to include terms associated with the large eigenvalues in the exponential sums.  Recall that the time-evolution operator of the Lanczos propagator is:
\begin{equation}
\hat{\textbf{U}}^L= \sum_k \ket{\lambda_k}\exp(-i \lambda_k \delta t)\bra{\lambda_k}\,. 
\end{equation}
The eigen-pairs $\lambda_k$ and $\ket{\lambda_k}$ change with each Lanczos iteration, $k$. 
In contrast to what one might expect, the tests show that the convergence of this operator is actually more dependent on the accuracy of the  
lower-lying eigen-pairs rather than the high-lying ones. 
  Indeed our tests show that when the wave-function convergence is achieved, for a given $k$ number of iterations , the terms associated with higher eigenvalues in the exponential sum above have negligible effect on the convergence of $\hat{\boldsymbol{U^L}}$. This strongly suggests that other strategies employing a larger initial vector space, such as block Lanczos~\cite{cullum_block_1974} or Davidson~\cite{davidson_iterative_1975}, might be profitably employed to increase efficiency.

\subsection{\label{subsec:ati}Above Threshold Ionization Spectrum}
If the laser field is sufficiently strong, the electron may be ionized by the absorption of more than one photon of light, even if the fundamental laser frequency is insufficient to photoionize the electron. In addition, as the ionized electron moves away from its ionic core, it may absorb additional photons, elevating its kinetic energy.  This process, known as above threshold ionization, can lead to other interesting physical effects.  The interested reader should consult the references for more details \cite{eberly_above-threshold_1991,schwengelbeck_ionization_1994,bucksbaum_introduction_1990}.

Once the electron is freed from the binding energy of the core, it moves away from the center with a velocity  = - $E_0 \cos(\omega t)  /\omega$ \footnote[8]{remember that $F=qE$, but in atomic units $q$=1 for an electron. Newton's second law states $F=ma$, but again in atomic units $m$=1. Therefore $\int a(t)dt= v(t) = -(E_0/\omega)\cos \omega t$}. This oscillating velocity is called the quiver velocity and is fundamental to the motion of a charged particle in an electric field.  As shown by Keldysh~\cite{Keldysh}, the square root of the ratio of the ionization potential to the  average kinetic energy of the electron in the field can be used to estimate when the motion in the field may be treated as a perturbation or not.

Experiments with high-intensity photoionization of atoms have verified these basic ideas.  They show a sequence of peaks in the electron energy spectrum that are separated by photon energies that are related to $\omega$; the hallmark of multi-photon ionization~\cite{agostini_free-free_1979,corkum_above-threshold_1989}.

As a final test of the propagation methods discussed in the previous sections, we reproduce some of the results of Javanainen et al.\cite{j._javanainen_numerical_1988} for the above threshold ionization (ATI) spectrum. As a consequence of the spatial discretization inherent in the numerical method, the authors of \cite{j._javanainen_numerical_1988} take the approach of finding probabilities of the photo-electron spectrum by averaging over the odd and even eigenstates of the atom.

Using this approach the photo-electron spectrum probability is defined as,
\begin{equation}
\label{eq:ati_probability_spectrum}
P(E_{1/4}) = \frac{|\braket{\phi_k | \psi}|^2}{E_{k+1}-E_{k-1}}+\frac{|\braket{\phi_{k+1} | \psi}|^2}{E_{k+2}-E_{k}}
\end{equation}
where 
\begin{equation}
\label{eq:e_1/4}
E_{1/4}=\frac{1}{4}(E_{k-1}+E_k+E_{k+1}+E_{k+2}).
\end{equation}
Here $k=0,1,...$, $\phi_k$ and $E_k$ are the $k^{th}$ eigenstate and eigenvalue of the undisturbed atom, $H_0$.

\begin{figure}[ht]
 \centering
 \includegraphics[scale=0.36]{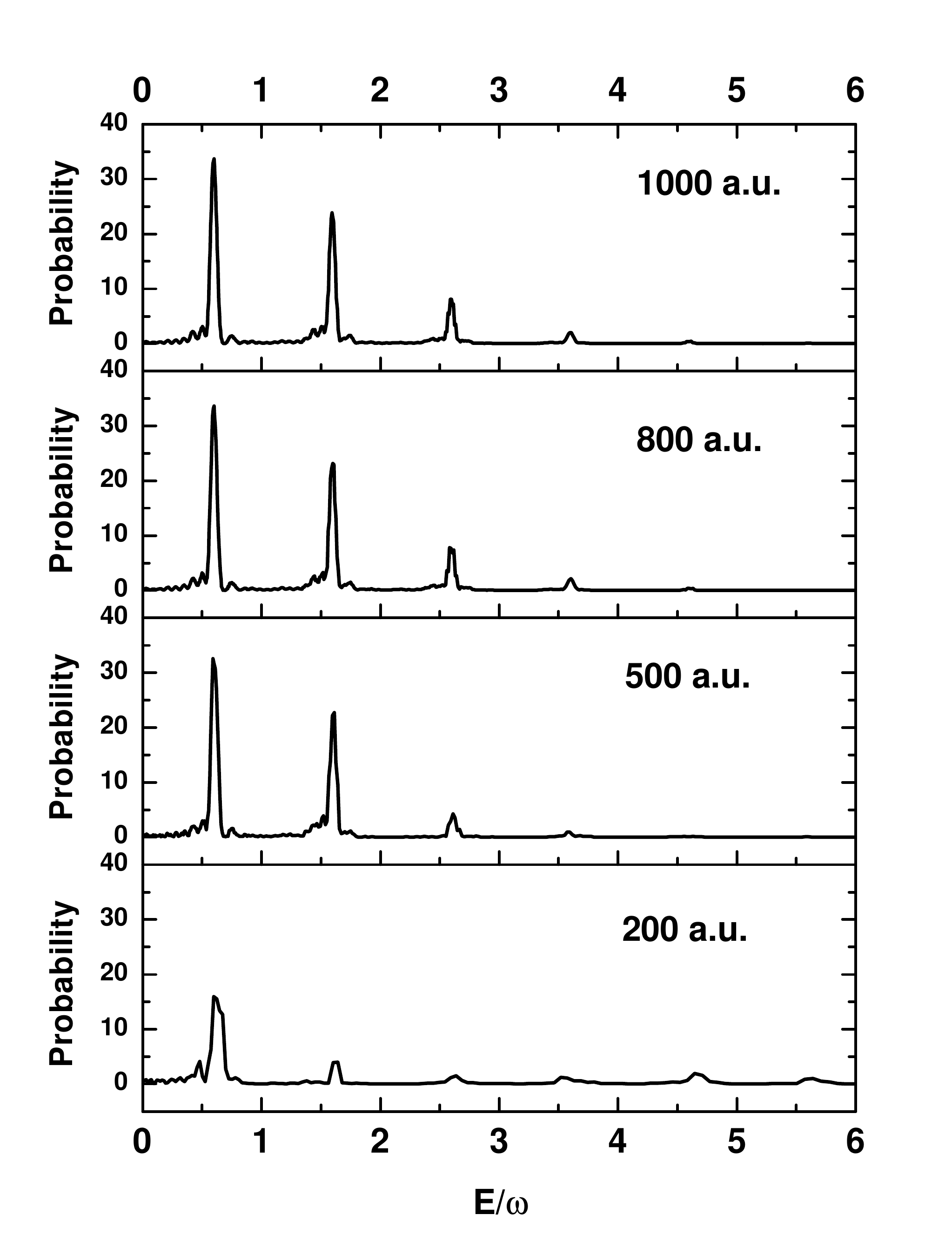} \caption{Graphs show the convergence of the ionization spectrum of our model problem with respect to size of the computational region using the Crank-Nicolson method.  The results are for a square pulse with $E_0$=0.1, $\omega$=0.148 and at a pulse length of 16+1/4 cycles. The other approaches discussed in the paper give similar results as long as the time-steps used conform to Table \ref{table:timingcompare}.  The results are independent of gauge.}
 \label{figure:populationgroundconverge}
\end{figure}

As can be seen in Fig. \ref{figure:populationgroundconverge}, convergence of the ionization probabilities can depend on the size of the computational region and the energy of electrons ionized.  This is not too surprising since the true continuum states do not vanish at the boundaries of the box and the higher the energy, the larger the box size needed for convergence. Here a box of 800 a.u., or larger, on either side of the origin produces converged spectra.

If one defines the probabilities as in~Eq.\eqref{eq:e_1/4}, the result depends critically on whether the wavefunction is in the length or velocity gauge. In the length gauge, the quiver motion of the electron is generally non-zero except at times $t=n+1/4$ or $n+3/4$ of the laser cycle.  If the spectra are computed at other times, they show different features. In the velocity gauge, the quiver motion is always zero so the spectra are stationary.  

It is legitimate to ask which gauge is ``correct", however, the widely adopted answer (which we use) is to perform the computation in either gauge and convert back to the length gauge before computing the probabilities.   But, we always compute those probabilities at that point in the laser cycle where the quiver velocity is zero.  In reality, the measurement of the probabilities should be made by propagating the wavepacket a few cycles after the pulse has been turned off.  If that is done, the issue of the quiver velocity is irrelevant.  

In one computational experiment we examined the probabilities when the quiver motion was non-zero.  The quiver velocity adds kinetic energy to the electrons moving to the right and to the left of the origin. This results in a doubling of the peaks at times when quiver velocity is not exactly zero. Any method which employs a variable $\delta t$ needs to ensure at the end of the propagation when the probabilities are measured that the quiver velocity is zero.

\section{\label{sec:conclusions} Conclusions} 
We presented a study of the soft-core, one-dimensional hydrogen atom in a strong electromagnetic field~\cite{j._javanainen_numerical_1988,eberly_high-order_1989}.  This model has been shown to be a reasonable approximation of the full hydrogen atom but does not include any effects due to angular momentum coupling.  The solution was propagated in time using a number of methods; Crank-Nicholson~\cite{j._crank_practical_1947}, various flavors of Split Operator~\cite{feit_solution_1982,de_raedt_product_1987,hermann_split-operator_1988,suzuki_fractal_1990,hochbruck_krylov_1997,Schneider2011,gonoskov_single-step_2016,jiang_efficient_2017}, the short iterative Lanczos~\cite{cornelius_lanczos_iteration_1950,paige_computation_1971,cullum_lanczos_1985,tae_jun_park_unitary_1986}, Chebychev propagator~\cite{talezer_accurate_1984}, and a combination of Lanczos and split operator.  The study examined the effects of the temporal and spatial step sizes on the efficiency of the propagation method as well as the question of using the length vs. velocity gauge.  For this study, both gauges produced identical results.  In more realistic problems, the velocity gauge might be more efficient for calculations requiring propgation to very large distances.~\cite{j._javanainen_numerical_1988}. 

Our conclusions demonstrate the efficiency of a method can depend heavily on the spatial step size and how one propagates in time.
In addition, and counterintuitive to established lore, we found that explicit methods can be used with larger time-steps than implicit ones, as long as the spatial grid is not too fine.

Section~\ref{sec:compmethods} outlined performance of each technique. As seen in Table~\ref{table:timingcompare}, Crank-Nicolson excels in computation involving small $\delta t$. The differences between convergence of CN and Lanczos were displayed in Fig~\ref{figure:CloseUp}. Additional computational experiments proved the flexibility of Lanzcos and Chebychev iterations in various sizes of spatial grid combined with Split-Operator propagation. 
  
  Table~\ref{table:timesfixeddtbiggerbox} indicates that other methods (block Lanczos, Davidson, etc.) may further improve computation time and accuracy given the convergence of Split-Operator+Lanczos method. The results also match the above-threshold ionization of Javanainen et al.~\cite{j._javanainen_numerical_1988}.
  
\section*{Acknowledgment}  
  We thank Klaus Bartschat of Drake University, Luca Argenti from the University of Central Florida and Lee Collins of Los Alamos National Laboratory for many helpful discussions and comments which greatly improved the manuscript. ML and HS thank the NIST's SURF program for supporting them during this project. 
  
\appendix

\section{\label{appsec:TDSE}Time Dependent Schr{\"o}dinger Equation}
  The eigenkets of a quantum mechanical system at two different times, $t$ and $t^\prime$, are related to each other through a time-evolution operator, $\hat{\textbf{U}}$,~\cite{sakurai_modern_1993}
  \begin{equation}
   \label{appeq:timeevolution}
   \ket{\psi(t^\prime)} = \hat{\textbf{U}}(t^\prime;t) \ket{\psi(t)}\,.
  \end{equation}
  Since the eigenkets of the system are normalized at any time, the time-evolution operator is unitary,
  \begin{equation}
  \hat{\textbf{U}}^\dagger(t^\prime ; t)~\hat{\textbf{U}}(t^\prime ; t) = 1\,.
  \end{equation}
  In addition,
  \begin{equation}\label{appeq:compositionalrelation}
 \hat{\textbf{U}}(t_2 ; t_0) = \hat{\textbf{U}}(t_2;t_1)~\hat{\textbf{U}}(t_1 ; t_0) 
  \end{equation}
  It is clear that  time-evolution operator satisfies,
  \begin{equation}
  \lim_{t^\prime \rightarrow t}\hat{\textbf{U}}(t^\prime ; t ) = 1\,.
  \end{equation}
  For an infinitesimal $dt$, an operator $\hat{\textbf{U}}$, having all the above properties, satisfies, 
  \begin{equation}
  \hat{\textbf{U}}(t+dt; t) = 1 - i \Omega~dt\,,
  \end{equation}
  where $\Omega$ is any Hermitian operator having the dimension of inverse time. 
  Replacing $\Omega$ by $H$ yields,
  \begin{equation}\label{appeq:infinitesimalTE}
 \hat{\textbf{U}}(t+dt; t ) = 1 -  i \mathbf{H}~dt \,. 
  \end{equation}
  The time-dependent Schr{\"o}dinger Equation is then derived by using \eqref{appeq:infinitesimalTE} and  \eqref{appeq:compositionalrelation} where $t_2 \rightarrow t + dt$ and $t_1 \rightarrow t$:
  \begin{align}
  \hat{\textbf{U}}(t+dt ; t_0) &= \hat{\textbf{U}}(t+dt ; t)~\hat{\textbf{U}}(t;t_0)  \, \nonumber \\
                            &= \left( 1 -  i \mathbf{H}~dt \right) \hat{\textbf{U}}(t;t_0)\,.
  \end{align}
  \begin{equation}\label{appeq:TIMEEVOLTDSE}
  \hat{\textbf{U}}(t + dt ; t_0) - \hat{\textbf{U}}(t;t_0) = -  i \mathbf{H} ~dt ~\hat{\textbf{U}}(t;t_0)\,.
  \end{equation}
    \begin{equation}
    i \frac{\partial}{\partial t}~\hat{\textbf{U}}
  (t;t_0) = \mathbf{H}~\hat{\textbf{U}}(t;t_0)\,.
  \end{equation}
  Multiplying from left by $\ket{\psi(t_0)}$ yields,
    \begin{equation}\label{appeq:TDSE}
  i\frac{\partial}{\partial t}~\psi(t) = \mathbf{H}~\psi(t)\,.
  \end{equation}
  the time dependent {\Schro} equation.

\section{\label{appsec:SolutionsTDSE}Solutions to TDSE}
  \label{sec:solutions}
  There are three distinct solutions to the TDSE \eqref{appeq:TIMEEVOLTDSE} depending on the properties of the Hamiltonian involved~\cite{sakurai_modern_1993}:
  \begin{enumerate}
  \item If the Hamiltonian is time-independent the time-evolution operator is
  \begin{equation}
  \hat{\textbf{U}}(t;t_0) = \exp{ \left( -i \mathbf{H}(t - t_0)\right)}\,. 
  \end{equation}
  
  \item If the Hamiltonian is time-dependent, but the Hamiltonians at different times commute with each other,
  \begin{equation}
   \hat{\textbf{U}}(t;t_0) = \exp{ \left( -i \int^t_{t_0}  \mathbf{H}(t')~dt' \right)}\,. 
  \end{equation}
 \item If the Hamiltonian is time-dependent and the Hamiltonians at different times do not commute with each other,

 \begin{widetext}
  \begin{equation}
   \hat{\textbf{U}}(t;t_0) = 1 + \sum_{n=1} ^{\infty}  \left( -i\right)^n \int^t_{t_0} dt_1 \int^{t_1}_{t_0} dt_2 \cdots \int^{t_{n-1}}_{t_0} dt_n~\mathbf{H}(t_1) \mathbf{H}(t_2) \cdots \mathbf{H}(t_n)\,,
  \end{equation}
 
which is known as a Dyson series and is ordered in time. 
\end{widetext}   
  \end{enumerate}








\bibliography{main.bib}

\end{document}